\documentclass[fleqn,usenatbib]{mnras}

\usepackage{newtxtext,newtxmath}

\usepackage[T1]{fontenc}
\usepackage{ae,aecompl}

\usepackage{graphicx}	
\usepackage{amsmath}	
\usepackage{amssymb}	
\usepackage{subfigure}

\title[Microlens masses from Gaia]{Microlens mass determination for {\sl Gaia}'s predicted photometric events}

\author[P. McGill et al.]{Peter McGill,$^{1}$\thanks{E-mail: pm625@cam.ac.uk, lsmith,nwe,vasily@ast.cam.ac.uk}
Leigh C. Smith,$^{1,2}$
N. Wyn Evans,$^{1}$
Vasily Belokurov,$^{1,3}$
\newauthor Zenghua Zhang$^{4}$\thanks{PSL Fellow}\\
$^{1}$Institute of Astronomy, University of Cambridge, Madingley Rd, Cambridge CB3 0HA, UK\\
$^{2}$School of Physics, Astronomy and Mathematics, University of Hertfordshire, College Lane, Hatfield AL10 9AB, UK\\
$^{3}$Center for Computational Astrophysics, Flatiron Institute, 162 5th Avenue, New York, NY 10010, USA\\
$^{4}$GEPI, Observatoire de Paris, Universit{\'e} PSL, CNRS, 5 Place Jules Janssen, 92190 Meudon, France }
\date{Accepted XXX. Received YYY; in original form ZZZ}

\pubyear{2015}

\begin{document}
\label{firstpage}
\pagerange{\pageref{firstpage}--\pageref{lastpage}}
\maketitle

\begin{abstract}
We used {\it Gaia} Data Release 2 to search for upcoming photometric microlensing events, identifying two candidates with high amplification. In the case of candidate 1, a spectrum of the lens (l1) confirms it is a usdM3 subdwarf with mass $\approx 0.11 M_\odot$, while the event reaches maximum amplification of $20^{+20}_{-10}$ mmag on November 3rd 2019 ($\pm$1d).  For candidate 2, the lens (l2) is a metal-poor M dwarf with mass $\approx 0.38 M_\odot$ derived from SED fitting, and maximum amplification of $10^{+40}_{-10}$ mmag occurs on 
June 3rd 2019 ($\pm$4d). This permits a new algorithm for mass inference on the microlens. Given the predicted time, the photometric lightcurve of these events can be densely sampled by ground-based telescopes. The lightcurve is a function of the unknown lens mass, together with 8 other parameters for all of which {\sl Gaia} provides measurements and uncertainties. Leveraging this prior information on the source and lens provided by {\it Gaia}'s astrometric solution, and assuming that a ground-based campaign can provide 50 measurements at mmag precision, we show for example that the mass of l1 can be recovered to within 20 per cent (68 per cent confidence limit).
\end{abstract}




\section{Introduction}

The advent of data from the {\sl Gaia} satellite has caused a flurry of interest in the prediction of future microlensing events, whether astrometric or photometric. Given precise stellar positions, parallaxes  and proper motions, it is a straightforward task to estimate whether a background object lies within the estimated Einstein radius of a foreground lens. The {\sl Gaia} data releases have proved to be a treasure trove for finding such events. First, using data from {\sl Gaia} Data Release 1 and the Sloan Digital Sky Survey, \citet{McG18} reported on a predicted astrometric microlensing event caused by the white dwarf LAWD 37 (WD 1142-645). Subsequent to Gaia Data Release 2~\citep{GaiaDR2}, two ongoing astrometric microlensing events were identified by \citet{Kl18a}. Next a systematic search for microlensing events was carried out by \citet[][herafter B18]{Br18}, who rediscovered the astrometric events caused by LAWD 37 and Stein 2015B, reported earlier by \citet{Sahu2017}. He also identified 9 further events which may exhibit detectable photometric and astrometric signatures. \cite{BN18} then extended B18's work and presented an almanac of 2,509 predicted microlensing events with closest approaches within the next century. Finally, \citet[][herafter K18]{Kl18} presented their systematic search, with different astrometry and event detectability cuts to B18. There have also been studies of predicted astrometric lensing events with pulsars~\citep{Of18}, and photometric events by nearby stars potentially hosting exoplanets ~\citep{Mu18}.

Here, we discuss two photometric events found in our own search through {\sl Gaia} Data Release 2 (DR2) from a different viewpoint. It is well-known that the photometric lightcurve of a microlensing event is degenerate and the mass of the lens cannot be extracted unless further information is available~\citep[e.g.,][]{An02}. By itself, the microlensing lightcurve for a point source lensed by a point mass provides a constraint only on the degenerate combination of mass and lens-source relative proper motion and parallax. However, if both lens and source are present in {\sl Gaia} DR2, then there are measurements for all these quantities with errors. This suggests a new method for mass measurement of predicted photometric microlensing events with {\sl Gaia}. If the lightcurve is densely sampled, as is possible for predicted events, then the extraction of the mass is a straightforward Bayesian inference problem using the astrometric quantities from {\sl Gaia} Data Release 2.

The paper is organised as follows. We review the basics of microlensing
in Section 2, and present our parametrisation and model of the photometric signal in Section 3.  We then identify suitable photometric candidates from {\sl Gaia} DR2 in Section 4, and use Monte Carlo simulations to demonstrate the efficiency of our new method for measuring the masses of microlenses in Section 5. Finally we highlight further implications of our work.

\begin{figure*}
\subfigure[]{
\includegraphics[width=\columnwidth]{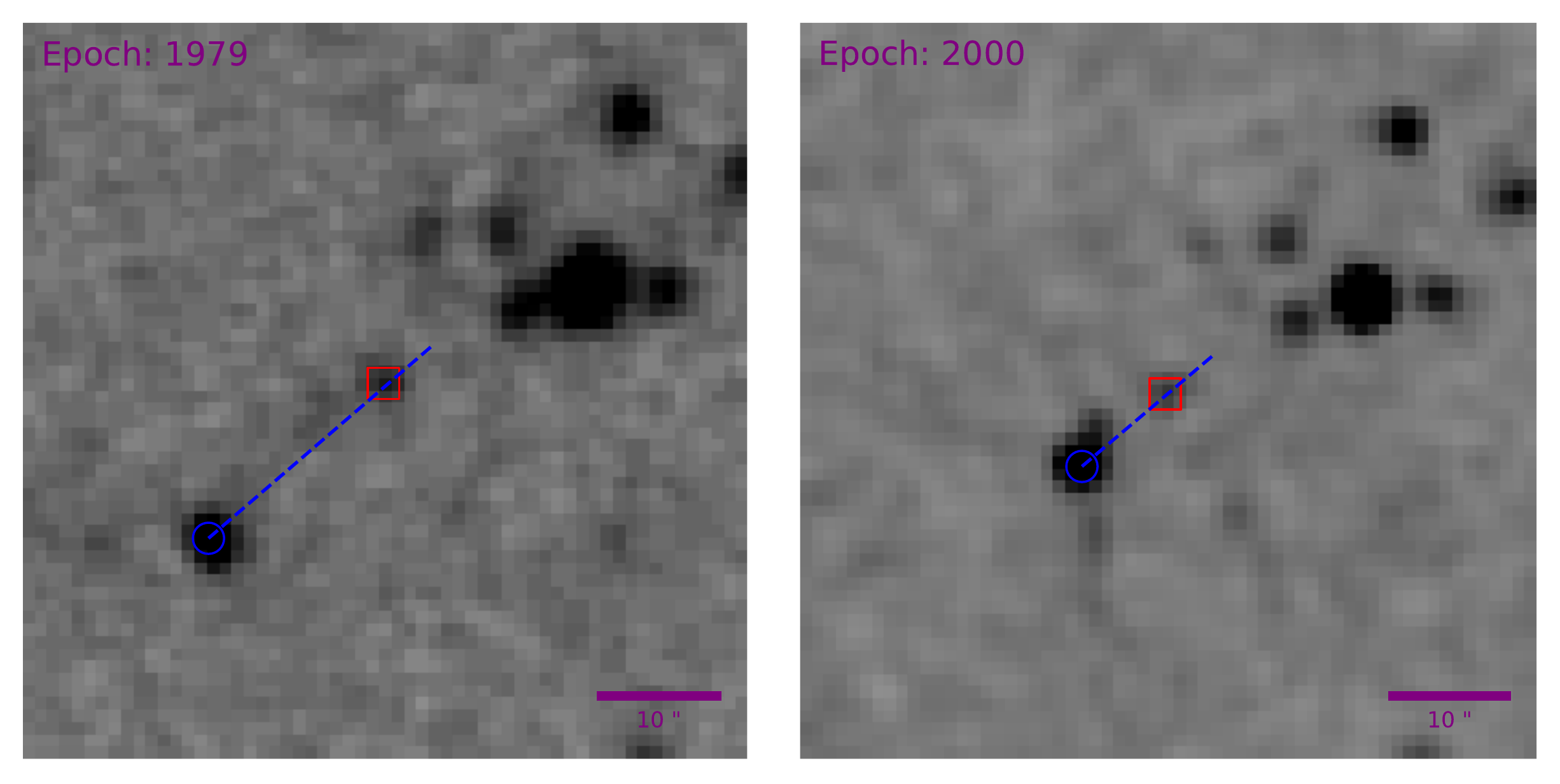}}
\subfigure[]{
\includegraphics[width=\columnwidth]{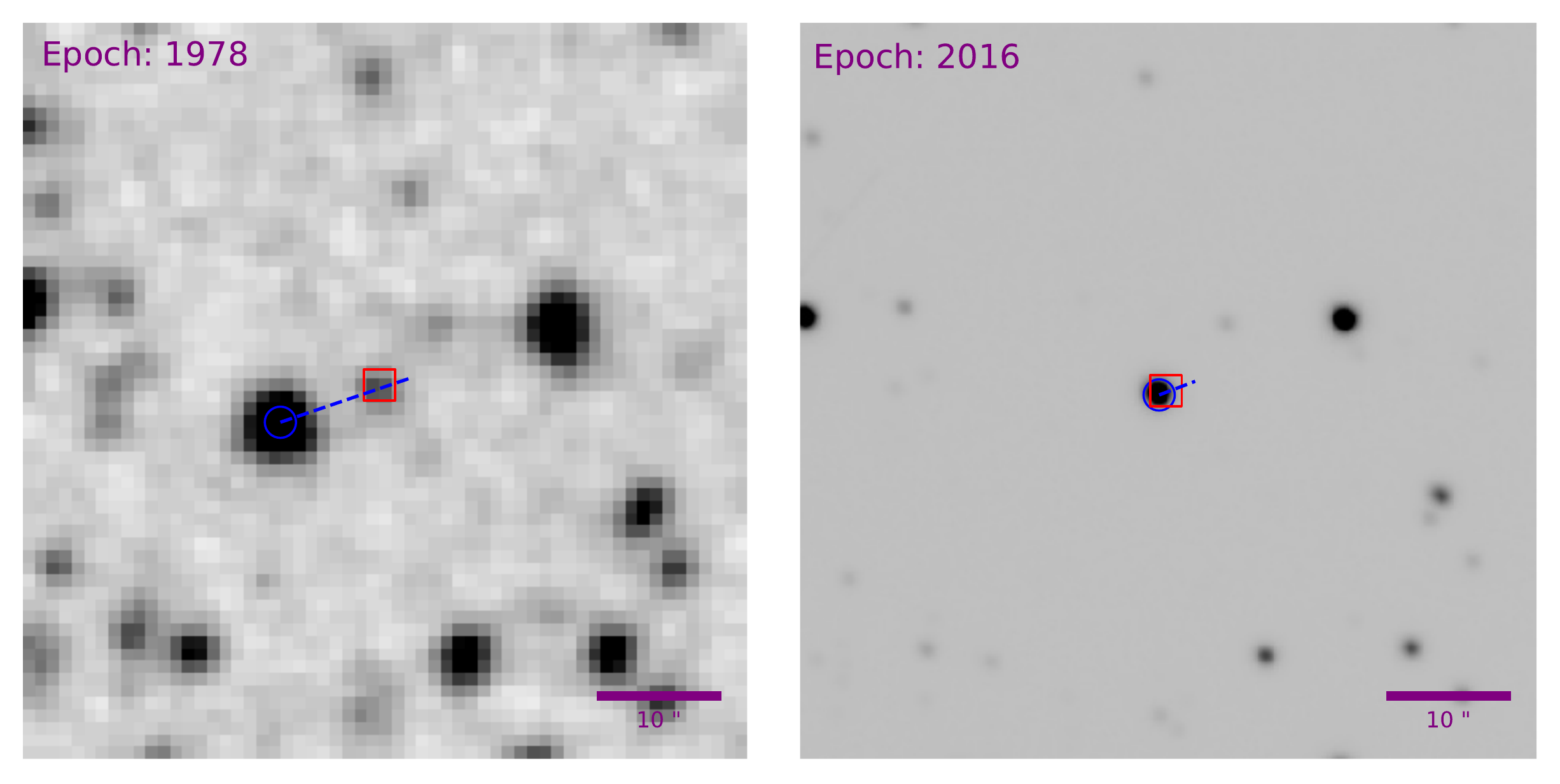}}
\caption{Stellar Field around each of the candidate events. (a) shows the stellar field around candidate 1. Left shows a Digital Sky Survey (DSS) image at epoch 1978. Right shows a 2MASS image at epoch 2000 \citep{2006AJ....131.1163S}. (b) shows the stellar field around candidate 2 at two different epochs. Left shows a DSS image at epoch 1978. Right shows a Dark Energy Camera Plane Survey (DECAPS) at epoch 2016 \citep{2018ApJS..234...39S}.In each image the positions of the lens and source at the image epoch
are indicated with a blue circle and red square receptively. The projected lens trajectory along its proper motion vector in shown with a blue dashed line in each case.}
\label{fig:stellarFeild}
\end{figure*}

\section{The Prediction of Microlensing Events}

\subsection{General Microlensing Principles}

Consider a point-like foreground object (lens) with mass $M_{l}$ and distance point-like background source. Microlensing occurs when a massive lens intervenes in the line of sight between an observer and a distance background source. When the lens and source have a non-zero angular separation $\Delta\phi$, a bright major ($+$) and fainter minor ($-$) image of the source are formed. The positions of the images along the line of the lens-source separation are~\citep{Paczynski1986} 
\begin{equation}
\frac{\theta_{\pm}(u)}{\text{mas}} = \frac{1}{2}\left[\pm\sqrt{u^{2}+4}+u\right]\frac{\Theta_{E}}{\text{mas}}.
\end{equation}
Here $\Theta_{E}$ is the angular Einstein radius and is given by
\begin{equation}
\frac{\Theta_{E}}{\text{mas}}  =  90.2 \text{ } \sqrt{\frac{M_{l}}{M_{\sun}}\left(\frac{\text{pc}}{D_{L}}-\frac{\text{pc}}{D_{s}}\right)}
 =  2.85 \text{ } \sqrt{\frac{M_{l}}{M_{\sun}}\frac{\varpi_{\text{rel}}}{\text{mas}}}
\end{equation}
where $u=|\Delta\phi|/\Theta_{E}$ is the normalised lens-source separation and $\varpi_{\text{rel}} = \varpi_{l}-\varpi_{s}$ is the relative lens source parallax. The amplifications of the two source images are~\citep{Paczynski1986}
\begin{equation}
A_{\pm}(u) = \frac{u^{2}+2}{2u \left(u^{2}+4\right)^{1/2}}\pm1.
\end{equation}
As the lens and source approach one another and the microlensing event unfolds, there are two observable effects. In the case of an unresolved luminous lens and source system, an apparent brightening of the lens-source blend is observed (photometric microlensing). The observed flux of the blend in a band and location denoted by the index $k$ is (assuming the only contaminating flux is from the lens), 
\begin{equation}
F^{k}_{\text{obs}}(u) = F^{k}_{s}A(u) +F^{k}_{l},
\label{eq:photSig}
\end{equation}
where $F^k_{l}$ is the flux of the lens, $F^k_{s}$ is the unlensed flux of the source, and $A(u)=A_{+}(u) + A_{-}(u)$ is the amplification from both the major and minor images. $F^{k}_{\text{obs}}$ is maximal when the lens and source are at closest approach ($u=u_{\text{min}}=|\Delta\phi_{\text{min}}|/\Theta_{E}$). The peak photometric amplification for the lens-source blend is then,

\begin{equation}
A_{\text{max}} = \frac{F^{k}_{\text{obs}}(u_{\text{min}})}{F^{k}_{s}+F^{k}_{l}} = 2.5 \log\left[\frac{F^{k}_{\text{obs}}(u_{\text{min}})}{F^{k}_{s}+F^{k}_{l}}{}\right] \quad \text{mag}
\end{equation}

The timescale for photometric events is the Einstien time and given by the time taken to cross the Einstein radius, $t_{E} =\Theta_{E}/\mu_{\text{rel}}$, where $u_{\text{rel}}$ is the relative lens-source motion. The photometric amplification is only significant for lens-source separations of the same order or less than the Einstein radius ($u\sim 1$). For a $\sim 0.5 M_{\sun}$ lens at $50$ pc this correspond to closest approach $\Delta\phi_{\text{min}}\sim 10$\, mas.

Additionally, the source images cause an apparent excursion of the lens-source light centroid position (astrometric microlensing). This effect can unfold in two regimes. When the system is completely unresolved and (i.e the lens and source images are blended) the apparent astrometic shift due to microlensing is given by \cite{Dominik2000} as

\begin{equation}
\frac{\delta(u)}{\text{mas}} = \frac{u}{1+f_{l}/f_{s}}\left[\frac{u^{2}+3}{u^{2}+2+(f_{l}/f_{s})u\sqrt{u^{2}+4}}\right]\frac{\Theta_{E}}{\text{mas}}.
\end{equation}

Typically this shift is maximal ($\delta_{\text{max}}$) when $u=u_{\text{min}}$. In the case when the source and lens are fully resolved, we observe an apparent astrometric shift of the source from its unlensed position due to the position of the major source image given by \cite{Sahu2017} as,

\begin{equation} 
\frac{\delta_{+}(u)}{\text{mas}} = \frac{1}{2}\left[\left(u^{2}+4\right)^{1/2}-u\right]\frac{\Theta_{E}}{\text{mas}}.
\label{eq:centroidshift}
\end{equation}

This shift is maximal when $u=u_{\text{min}}$. {\it Gaia} has a source resolving limit of potentially  $\sim103$\,mas \citep{F16}. It is therefore possible for astrometric microlensing events as seen by {\it Gaia} to unfold in both regimes. For events with small enough closest approach, an astrometric shift according to $\delta_{+}$ is observed until the lens image system becomes unresolved. We then see an astrometric shift suppressed by light from the lens $\delta$, as the lens and source reach closest approach. In the the case of an event $\Delta\phi_{\text{min}}< 103$\,mas, the maximum resolved shift seen by {\it Gaia}  will be $\delta_{\text{r,max}}=\delta_{+}(103 \text{ mas} / \Theta_{E})$.

\subsection{Predicted Motion of Stars}

If positions $(\alpha_{0},\delta_{0})$ at a reference epoch $t_{0}$, proper motions $(\mu_{\alpha *},\mu_{\delta})$ and parallaxes ($\varpi$) of an object are known, we may compute the expected position of the object on the celestial sphere at time t.$\bphi(t)=(\alpha(t),\delta(t))$, as seen by an observer on Earth. There is both a contribution to the object's motion from its proper motion vector $\bmu$ and parallax $\mathbfit{P}(t)$
\begin{align}
\bphi(t) &\approx  \bphi_{0} + 
\bmu \left[ t- t_{0} \right] + \varpi\mathbfit{P}(t) \\
&=
\begin{pmatrix}
\alpha_{0} \\
\delta_{0}
\end{pmatrix}
+ [t-t_{0}]
\begin{pmatrix}
\mu_{\alpha *} / \cos\delta_{\text{ref}}\\
\mu_{\delta} 
\end{pmatrix} + \varpi\mathbfit{P}(t)    
\label{eq:motion}
\end{align}
with
\begin{equation}
\mathbfit{P}(t) = 
\begin{pmatrix}
\left[\sin\alpha_{\text{ref}}X(t)-\cos\alpha_{\text{ref}}Y(t)\right]/\cos\delta_{\text{ref}}\\
\cos\alpha_{\text{ref}}\sin\delta_{\text{ref}}X(t)+\sin\alpha_{\text{ref}}\sin\delta_{\text{ref}}Y(t)-\cos\delta_{\text{ref}}Z(t)
\end{pmatrix}
\end{equation}
where $X(t)$, $Y(t)$ and $Z(t)$ are the Cartesian solar system barycentric coordinates in AU of the earth on the ICRF at time t. They were computed from NASA JPL's Horizons Ephemeris and retrieved via the {\tt astropy} python package \citep{2018arXiv180102634T}. ($\delta_{\text{ref}},\alpha_{\text{ref}}$) is the reference position and usually taken to be equal to the initial position $(\alpha_{0},\delta_{0}).$ For DR2 the reference time $t_{0}$ is 2015.5 Julian Years.  In this study we handle sources with negative or no measurement of parallax as having a parallax equal to zero. Eq \ref{eq:motion} allows the calculation of the on sky separation $\Delta\phi$ of a lens and source, and combined with a mass estimate for lens, allows the prediction of future microlensing events.

\label{sec:model}
\section{A model of the Photometric Signal}

In order to model the photometric signal of a microlensing event, we choose a parameterisation of the light curve that naturally allows prior knowledge of astrometric quantities derived from DR2 data on the source and lens to be used easily. 

We fix the reference position for both the lens and source at the position of the source $\alpha_{0s},\delta_{0s}$. Using equation (\ref{eq:motion}) this allows us to write the angular separation of a source (s) and lens (l) at time $t$ in terms of their relative astrometric quantities,
\begin{align}
\Delta\Phi \approx& 
\begin{pmatrix}
\alpha_{0l}-\alpha_{0s} \\
\delta_{0l}-\delta_{0s}
\end{pmatrix}
+
[t-t_{0}]
\begin{pmatrix}
[\mu_{\alpha*l}-\mu_{\alpha*s}] / \cos\delta_{0s} \\
\mu_{\delta l} - \mu_{\delta s}
\end{pmatrix}
\\
+&
[\varpi_{l}-\varpi_{s}]\textbf{P}(t)
\\
=&
\begin{pmatrix}
\alpha_{0,\text{rel}} \\
\delta_{0,\text{rel}}
\end{pmatrix}
+
[t-t_{0}]
\begin{pmatrix}
\mu_{\alpha*,\text{rel}} / \cos\delta_{0s} \\
\mu_{\delta*,\text{rel}}
\end{pmatrix}
+
\varpi_{\text{rel}}\textbf{P}(t).
\end{align}
It is now possible to write a model for the photometric microlensing signal parameterised by relative astrometric quantities and base fluxes we have prior knowledge of from DR2 for both the source and lens. Rewriting eq.~(\ref{eq:photSig}) as a function of time with its explicit parameters dependencies, we obtain our model for the photometric signal as
\begin{equation}
F^{k}_{\text{obs}}\left(t_{i}; \mathbfit{p},F^{k}_{s},F^{k}_{l}\right) = F^{k}_{s} A(t_{i};\mathbfit{p}) + F^{k}_{l} + \epsilon_{i}.
\label{eq:model}
\end{equation}
where we assume $\epsilon_{i}$ is Gaussian noise. Here $\mathbfit{p}$ is the vector of parameters that control the amplification and hence the shape of the microlensing light curve and is,
\begin{equation}
\mathbfit{p} = \{\alpha_{0,\text{rel}},\delta_{0,\text{rel}},\mu_{\alpha*,\text{rel}},\mu_{\delta,\text{rel}},\varpi_{\text{rel}},M_{l}\}.
\end{equation}
Our task is fit the light curve model outlined in eq~(\ref{eq:model}) to a set of $N$ observed photometric data points $\mathcal{D}^k = \{t_{i},f^{k}_{i},\sigma_{f,i}\}_{i=1}^{N}$, where $f^{k}_{i}$ is the observed flux of the lens-source blend at time $t_{i}$ with measurement variance $\sigma^{2}_{f,i}$. Under the assumption that noise terms $\epsilon_{i}$ are independent, we may write the likelihood function of the microlensing light curve as,
\begin{equation}
\mathcal{L}(\mathcal{D}|\mathbfit{p},F^{k}_{s},F^{k}_{l}) =  \displaystyle\prod_{i=1}^{N} \mathcal{N}\left[f^{k}_{i} | F^{k}_{\text{obs}}\left(t_{i}; \mathbfit{p},F^{k}_{s},F^{k}_{l}\right),\sigma^{2}_{f,i}\right],
\end{equation}
where $\mathcal{N}(x|u_{x},\sigma^{2}_{x})$ is the normal density in the random variable $x$, with mean $\mu_{x}$ and variance $\sigma^{2}_{x}$. Using Bayes theorem, we write the posterior distribution or the probability of the model parameters given the data,
\begin{equation}
P\left(\mathbfit{p},F^{k}_{s},F^{k}_{l} |\mathcal{D}\right) \propto \mathcal{L}\left(\mathcal{D}|\mathbfit{p},F^{k}_{s},F^{k}_{l}\right) \times \text{Pr}\left(\mathbfit{p},F^{k}_{s},F^{k}_{l}\right).
\label{eq:posteroir}
\end{equation}
Here, Pr denotes the prior on the model parameters. For all parameters apart from the mass of the lens $M_{l}$, if the lens and source are present in DR2 Gaussian priors can be derived for these quantities. Finally, we place a uninformative flat uniform prior on $M_{l}$ as our aim is to infer $M_{l}$ from the data (see Table 2 for details on the priors). Overall, using the light curve for a predicted microlensing event combined with prior knowledge of the source and lens from their {\it Gaia} DR2 astrometric solutions will allow the determination of the lens mass.

\section{Candidate Event Search}

In order to find predicted photometric microlensing events, we take a high proper motion ($>150 \text{ mas yr}^{-1}$) sample of 168,734 lens stars from the {\it Gaia} DR2 source catalogue. In order to try and remove spurious high proper motion objects from our sample, we make further photometric cuts. We take high proper motion objects with \textit{Gaia} G-band magnitude $G < 19$ to allow for visual confirmation of the candidate events using current image data, and all objects with a measured {\it Gaia} BP-RP color so photometric mass estimates may be obtained for the lens. These further photometric cuts leave a sample 136,791 lens stars. To narrow our search we cross match each star in our lens sample with all {\sl Gaia} DR2 sources within a radius of 10 times the proper motion of the lens. This leaves $\sim$ 10,000 lens source pairs which we investigate further. We search for events with a closest approach within the remaining {\sl Gaia} mission time which we assume to be 2018 to 2022, and look for events with a closest approach separation $< 10$ \,mas, which are likely to have a detectable photometric signal. We find two such events.

\begin{table}
\caption{Lens, background source and event data for the two candidates. Data for all objects are from the {\it Gaia} DR2 source catalogue. The coordinates ($\alpha,\delta$) are on the ICRF and at epoch 2015.5 Julian Years. $\text{flux}_{G}$ is the {\it Gaia} G band flux, $G_{\text{abs}}$ is the absolute G band magnitude calculated via $G_{\text{abs}} = G + 5 + 5\log_{10}(\varpi/1000)$. AEN Sig is the excess astrometric noise significance parameter provided in DR2.}
\begin{tabular}{llll}
\hline
 & Candidate 1 Lens & Candidate 2 Lens \\
\hline
DR2 id &  5840411363658156032 & 5862333044226605056 \\
$\alpha_{0l}$ (deg$\pm$mas)&196.460398500$\pm$0.07 &196.506734481$\pm$0.03 \\
$\delta_{0l}$ (deg$\pm$mas)&-72.300995137$\pm$0.06 &-63.532796061$\pm$0.03 \\
$\mu_{\alpha*l}$ (mas/yr) &362.72$\pm$0.13 &-209.59$\pm$0.05 \\
$\mu_{\delta l}$ (mas/yr) &306.51$\pm$0.12 &79.60$\pm$0.05 \\
$\varpi_{l}$ (mas) & 9.52$\pm$0.08 &6.71$\pm$0.04\\
$G$ (mag) &17.18 &15.18  \\
$\text{flux}_{G}$ (e-/s)  & 2500$\pm$1.95 &16000$\pm$12.67 \\
BP-RP (mag) &2.14 &2.16 \\
$G_{\text{abs}}$ & 12.07 & 9.31\\
AEN Sig &1.75 & 5.42\\
\hline
& Candidate 1 Source & Candidate 2 Source \\
\hline
DR2 id & 5840411359350016128 & 5862333048529855360 \\
$\alpha_{0s}$ (deg$\pm$mas)&196.459018289$\pm$0.13 & 196.506239773$\pm$0.36 \\
$\delta_{0s}$ (deg$\pm$mas)&-72.300626103$\pm$0.11  & -63.532706855$\pm$0.51\\
$\mu_{\alpha*s}$ (mas/yr) &-12.95$\pm$0.29 & -9.02$\pm$0.56\\
$\mu_{\delta s}$ (mas/yr) &2.92$\pm$0.23 & -4.42$\pm$0.91 \\
$\varpi_{s}$ (mas) &0.59$\pm$0.13  &-0.06$\pm$0.37 \\
$G$ (mag) &18.17 & 18.09\\
$\text{flux}_{G}$ & 1000$\pm$1.62 & 1100$\pm$8.89\\
BP-RP & 1.43 & - \\
AEN sig & 0 & 8.30\\
\hline 
& Candidate Event 1 & Candidate Event 2 \\
\hline
$t_{\text{min}}$ (Jyear) & 2019.839$\pm$0.003  & 2019.42$\pm$0.01 \\
$\Delta\phi_{\text{min}}$ (mas) & $5.83^{+1.26}_{-1.20}$ & $6.48^{+3.38}_{-3.36}$\\
$M_{l}$ ($M_{\sun}$) & $0.11\pm0.01$ & $0.38\pm0.06$ \\
$\Theta_{E} $ (mas) & $2.82\pm0.12$ & 4.56$^{+0.35}_{-0.37}$ \\
$u_{\text{min}}$ ($\Theta_{E}$) & $2.03^{+0.48}_{-0.47}$ & $1.41^{+0.78}_{-0.76}$\\
$t_{E}$ (d) & $2.2\pm0.1$ & $7.7\pm0.6$\\
$A_{\text{max}}$ (mmag) & $20^{+20}_{-10}$& $10^{+40}_{-10}$\\
$\delta_{\text{r,max}}$ (mas) & $0.077\pm0.007$ & $0.20^{+0.04}_{- 0.03}$ \\
$\delta_{\text{max}}$ (mas) & $0.569^{+0.010}_{-0.089}$& $0.04\pm0.01$\\
\end{tabular}
\label{tab:candevents}
\end{table}
\begin{table}
\caption{Form of the priors used for each parameter. $\mathcal{N}(\mu,\sigma^{2})$ is a normal distribution with mean $\mu$ and variance $\sigma^{2}$. $\mathcal{U}(a,b),\quad b>a$ a the uniform distribution and is 1 in the interval [a,b] and zero otherwise.}
\begin{tabular}{lll}
\hline
Parameter & Unit & Prior \\
\hline
$\alpha_{0,\text{rel}}$ & mas & $\mathcal{N}(\alpha_{0l}-\alpha_{0s},\sigma^{2}_{\alpha_{0l}} + \sigma^{2}_{\alpha_{0s}})$ \\
$\delta_{0,\text{rel}}$ & mas & $\mathcal{N}(\delta_{0l}-\delta_{0s},\sigma^{2}_{\delta_{0l}} + \sigma^{2}_{\delta_{0s}})$ \\
$\mu_{\alpha *,\text{rel}}$ & mas/yr & $\mathcal{N}(\mu_{\alpha*l}-\mu_{\alpha*s},\sigma^{2}_{\mu_{\alpha*l}} + \sigma^{2}_{\mu_{\alpha*s}})$ \\
$\mu_{\delta,\text{rel}}$ & mas/yr &  $\mathcal{N}(\mu_{\delta l}-\mu_{\delta s},\sigma^{2}_{\mu_{\delta l}} + \sigma^{2}_{\mu_{\delta s}})$\\
$\varpi_{\text{rel}}$ & mas & $\mathcal{N}(\varpi_{l} -\varpi_{s},\sigma^{2}_{\varpi_{l}}+\sigma^{2}_{\varpi_{s}})$\\
$F_{l}$ & ergs/s& $\mathcal{N}(F_{l},\sigma^{2}_{F_{l}})$ \\
$F_{s}$ & ergs/s& $\mathcal{N}(F_{s},\sigma^{2}_{F_{s}})$ \\
$M_{l}$ & $M_{\sun}$ & $\mathcal{U}(0,1)$ \\
\end{tabular}
\label{tab:priors}
\end{table}
\begin{figure}
\includegraphics[width=\columnwidth]{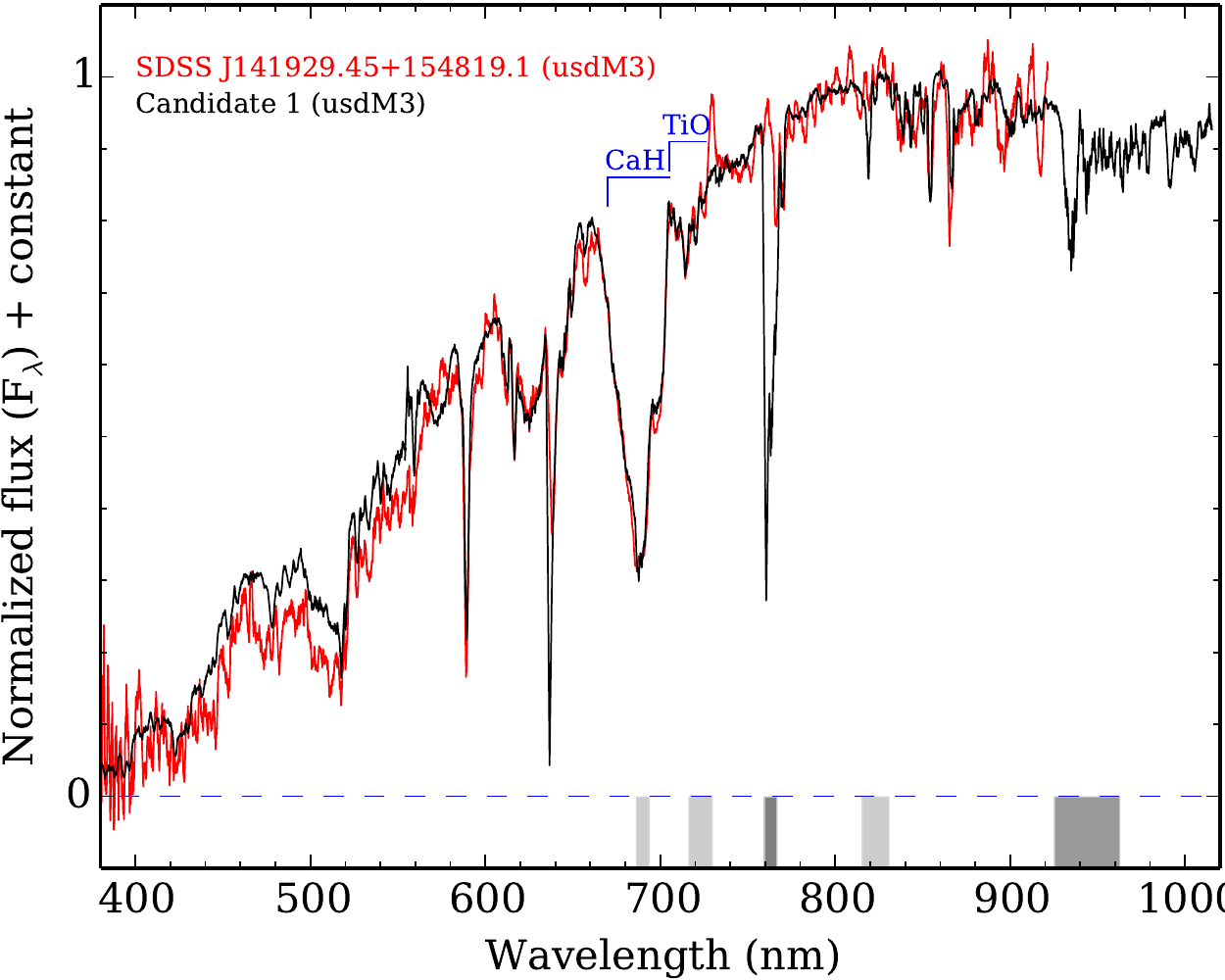}
\caption{The UVB and VIS spectrum of candidate 1 compared to a usdM3 subdwarf, SDSS J141929.45+154819.1. Telluric absorption regions are indicated with grey bands (not corrected). Lighter and thicker shaded bands indicate regions with weaker and stronger telluric effects.}
\label{fig:opspec}
\end{figure}
\begin{figure*}
\includegraphics[width=\textwidth]{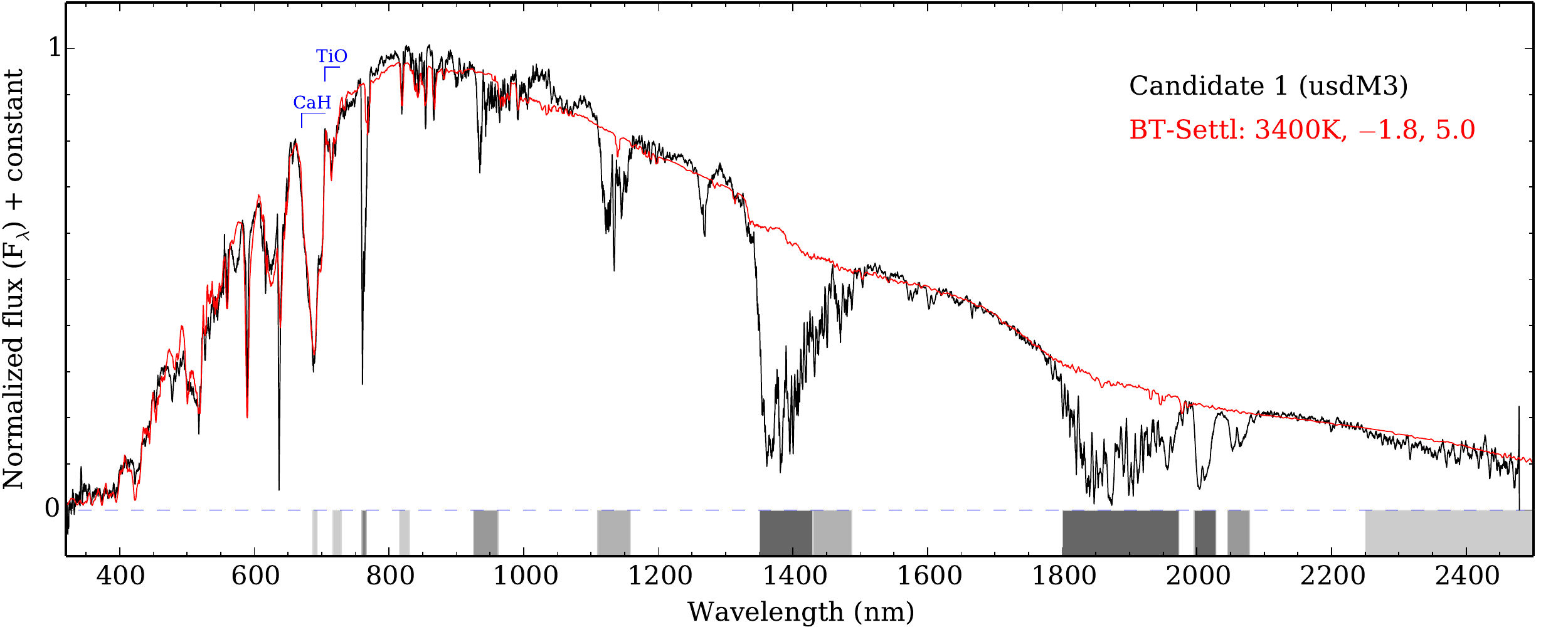}
\caption{The X-shooter spectrum of candidate 1 (black) compared to its best-fit BT-Settl model spectrum (red). Telluric absorption regions are indicated with grey bands (not corrected). Lighter and thicker shaded bands indicate regions with weaker and stronger telluric effects.}
\label{fig:nirspec}
\end{figure*}
\begin{figure*}
\includegraphics[width=\textwidth]{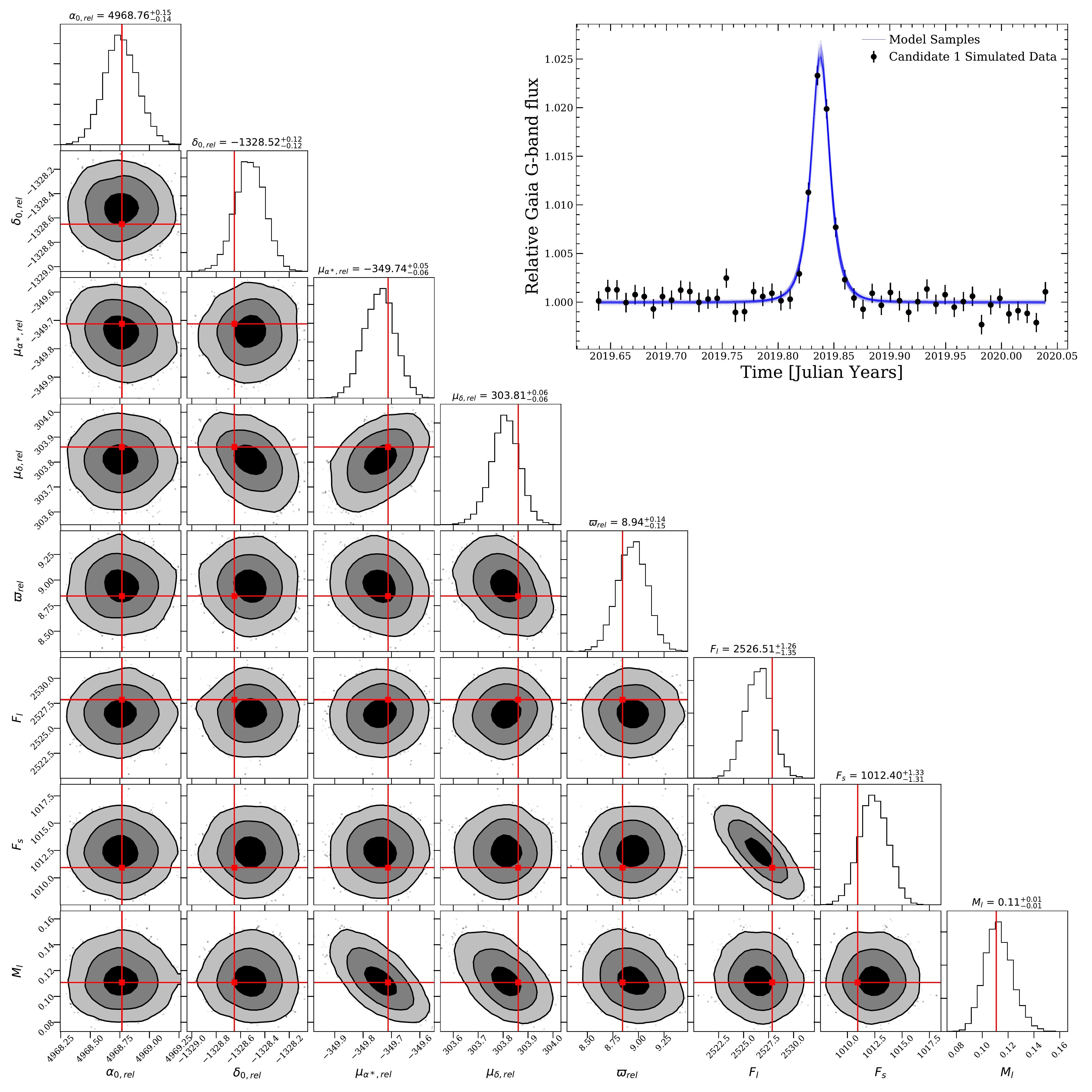}
\caption{Bottom Left: Marginal posterior distributions for the 8 parameters inferred from simulated microlensing light curve data for candidate 1. $\delta_{0,\text{rel}}$ and $\alpha_{0,\text{rel}}$ are int initial separations of the lens and source in units of mas. $\mu_{\alpha *,\text{rel}}$ and $\mu_{\delta,\text{rel}}$ are the relative lens source proper motion in units of mas/yr. $\varpi_{\text{rel}}$ is the relative lens-source parallax in units of mas. $F_{l}$ and $F_{s}$ are the lens and source fluxes in units of e-/s, and $M_{l}$ is the mass of the lens in units of $M_{\sun}$. Contours shows 1,2 and 3 $\sigma$ bands. Red lines indicate the truth values used to generate the data for each parameter. Top right: Back points indicated simulated date used for the inference with $\sim$ mmag photometric precision. Blue lines show samples of the corresponding inferred light curves.}
\label{fig:corner}
\end{figure*}
\begin{figure}
\includegraphics[width=\columnwidth]{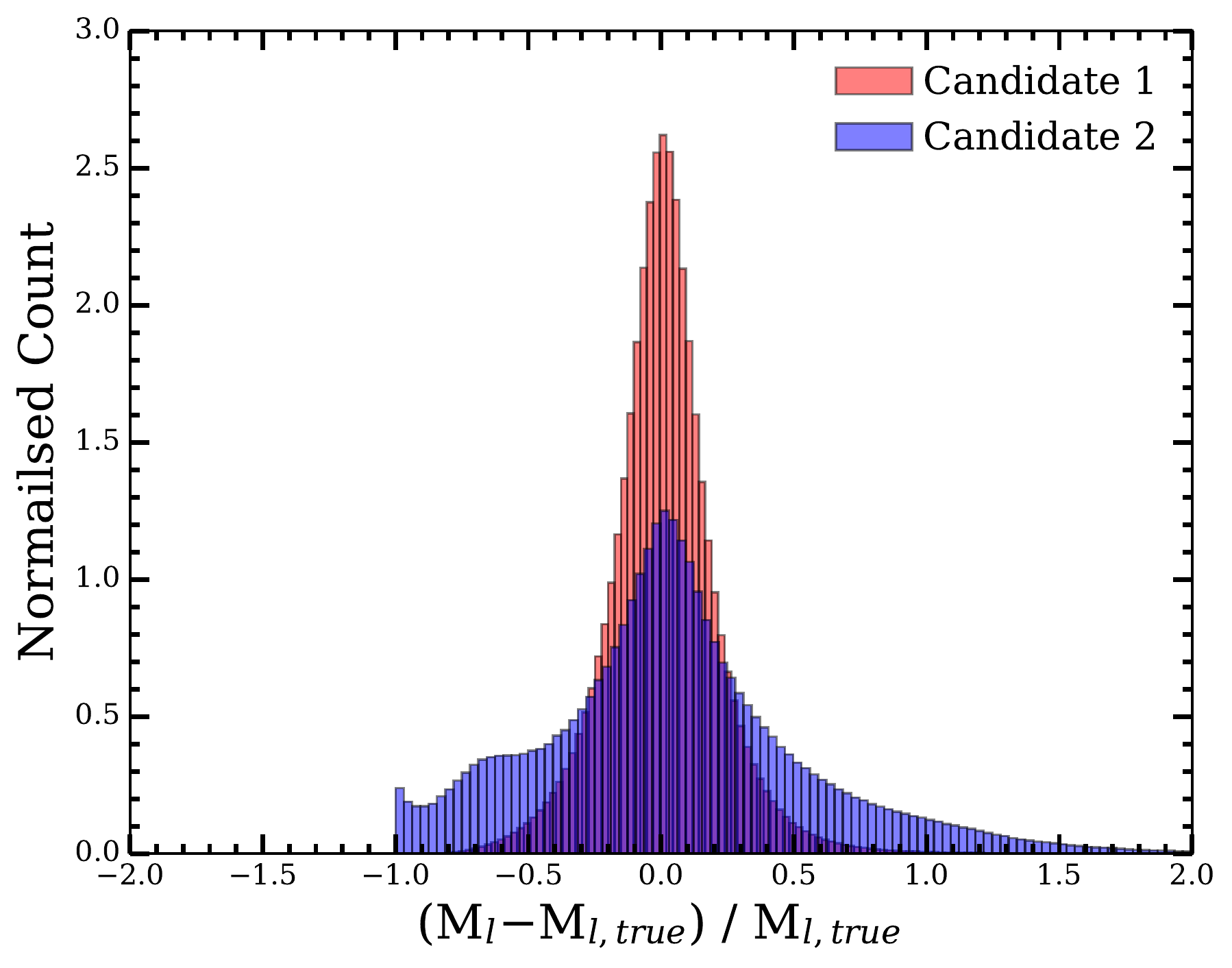}
\caption{Stacked normalised marginal mass samples from the 1000 simulated inferences for the two candidate events. For candidate 1, we recover the mass of the lens within $\pm20$ per cent with 68 per cent confidence. For candidate 2, the mass recovery is poorer at $\pm47$ per cent. The truncation in candidate 2's recovered mass distribution at -1.0 is due to the uniform prior lens mass prior (see table \ref{tab:priors}) forbidding unphysical negative lens masses.}
\label{fig:efficiency}
\end{figure}

\subsection{Candidate 1}

We predict that the lens with DR2 source id 5840411363658156032 (hereafter l1) will lens the light of a background star on November 3rd 2019 ($\pm$ 1d) or $2019.839$ $(\pm0.003)$ Julian Years, with a closest approach of $\Delta\phi_{\text{min}} =5.75^{+1.22}_{-1.16}$\,mas. Fig.~\ref{fig:stellarFeild} (a) shows the stellar field around l1 at two different epochs. Table \ref{tab:candevents}
contains information for both l1 and the background source. BP-RP and absolute G band magnitude $G_{\text{abs}}$ are consistent with l1 being a cool sub-dwarf.\\ 

We obtained the 300--2480 nm wavelength spectrum of l1 with the X-shooter \citep{ver11} on the Very Large Telescope on 2018 July 8 under a seeing of 0.61\arcsec ~and an average airmass of 1.61. The X-shooter spectrum was observed in an AB nodding mode with slits of 1.0\arcsec~ in the ultraviolet-blue (UVB), and 0.9\arcsec~ in the visible (VIS) and near-infrared (NIR) arms providing a resolving power of 5100, 8800 and 5100, respectively. The integration time was 2$\times$223 s in the UVB, 2$\times$235 s in the VIS, and 2$\times$250 s in the NIR. A wavelength and flux calibrated 2D spectrum of l1 was first reduced with European Southern Observatory (ESO) Reflex \citep{fre13}. Then we extracted a 1D spectrum from the 2D spectrum with {\scriptsize IRAF}\footnote{IRAF is distributed by the National Optical Observatory, which is operated by the Association of Universities for Research in Astronomy, Inc., under contract with the National Science Foundation.} task {\scriptsize APSUM}.

The original X-shooter spectrum of l1 has a signal-to-noise ratio (SNR) of 30 at 490 nm, 74 at 830 nm, 41 at 1200 nm, 39 at 1700 nm, and 52 at 2150 nm. Note that the original spectrum was smoothed  (using boxcar smooth with {\scriptsize IRAF SPLOT}) by 101 pixels in the UVB and VIS, and 51 pixels in the NIR for display in Figs \ref{fig:opspec} and \ref{fig:nirspec} which increased the SNR by about 10 and 7 times. 

Fig. \ref{fig:opspec} shows the UVB and VIS spectrum of l1. We classified it as an M3 ultra-subdwarf (usdM3), as it fits well to the usdM3 subdwarf, SDSS J141929.45+154819.1 observed by the Sloan Digital Sky Survey \citep[SDSS;][]{yor00}. The classification of M subdwarfs is based on the CaH and TiO absorption bands in the optical \citep{lep07}. The TiO absorption band gets weaker with the decrease of metallicity in M subdwarfs.  

As the CaH band is very sensitive to $T_{\rm eff}$ and the TiO band is very sensitive to [Fe/H], we therefore fitted our observed spectrum with BT-Settl model spectra \citep{all14} primarily by their CaH and TiO absorption bands and by the overall profile in a non-standard approach. We also applied linear interpolation between some models where this was able to improve the fit. We used a fixed gravity because early-type M subdwarfs have similar gravity around log $g$ = 5.0 according to model predictions. We gradually changed the  $T_{\rm eff}$ and [Fe/H] of models by steps of 50 K and 0.1 dex to find the best-fitting model by visual inspection focused on the CaH, TiO bands and overall profile. The best-fitting BT-Settl model for l1 has $T_{\rm eff}$ = 3400 K, [Fe/H] = $-$1.8 and log $g$ = 5.0. The atmospheric parameter uncertainties for l1 are around 100 K for $T_{\rm eff}$, 0.2 dex for [Fe/H] and 0.25 dex for log $g$. Fig. \ref{fig:nirspec} shows the full X-shooter spectrum of l1 compared to a best-fitting BT-Settl model spectrum. We placed l1 in a $T_{\rm eff}$ versus [Fe/H] space \citep[e.g. Figure 9 in][]{zha17} and found that it has a mass of 0.11$\pm$0.01 M$_{\sun}$ according to 10 Gyr iso-mass contours predicted by evolutionary models \citep{bar97,cha97}. 

Using the mass of l1 derived from atmospheric and evolutionary models, we estimate its Einstein radius $\Theta_{E}=2.82\pm0.12$\, mas. This corresponds to a peak photometric amplification of the lens source blend flux of $1.02^{+0.02}_{-0.01}$ or a change in magnitude or $20^{+20}_{-10}$ mmag. The Einstein time scale for this event is $2.2\pm0.1$ d. At the point l1 is still resolvable from the background source by {\it Gaia}, the maximum astrometric of the major source image is $0.077\pm0.007$\, mas. At closest approach, the maximum shift of the lens-source blend will be $0.569^{+0.010}_{-0.089}$\, mas.

This event was also independently discovered as candidate ME19 in B18. While we are in agreement with B18's closest approach and separation for this event, our predicted signal strengths differ considerably. The reason for this is that B18 classifies l1 as a late-type M dwarf (B18: sections 6.3,7), and adopts a mass of $0.25 M_{\sun}$, $\sim225$ per cent higher than our own. Consequently, B18 predicts significantly higher peak photometric and astrometric signals for this event. K18 also independently predicts this event as candidate \#4. Similarly to B18's predictions, we are in good agreement with K18's astrometric calculations for this event, however K18 adopts a mass of $0.17M_{\sun}$ for l1 (K18: table 2), $\sim$ 50 percent higher than ours. As a result K18 also predicts higher peak photometric and astrometric signals for this event.


\subsection{Candidate 2}

We also predict the lens l2 with {\it Gaia} DR2 source id 5862333044226605056 will lens the light of a background star with closest approach on June 3rd ($\pm$4d) or 2019.42$\pm$0.01 Julian Years, with closest approach of $6.48^{+3.38}_{-3.36}$ mas. Fig.~{\ref{fig:stellarFeild}} (b) shows the stellar field around l2 at two different epochs, and table \ref{tab:candevents} contains details on the lens, source and event. 

l2 is a known high proper motion object, visually confirmed by \citet{LS18} with VIRAC id 323066023. l2's BP-RP and $G_{\rm abs}$ are consistent with l2 being a mildly metal-poor M dwarf. Its tangential velocity hints at thick disk kinematics, further evidence for a somewhat low metallicity. In order to obtain a mass estimate for l2, we first fit {\it Gaia} BP, G, and RP, VISTA VVV survey J band, and WISE W1 and W2 photometry to model spectral energy distributions (SEDs) using the virtual observatory SED Analyizer \citep[VOSA;][]{VOSA08}. We omitted the VVV Z and Y band photometry due to relatively poor calibration in those bands, and H and K band photometry due to saturation. We note that flux from the background source will contaminate the WISE photometric measurements, but given its relative faintness contamination should be minor. We omitted the W3 and W4 upper limits from the SED fit but retained the W1 and W2 bands as the additional wavelength coverage offsets the negative impact of the potential contamination. We fit the photometry to the BT-Settl and BT-Dusty models \citep{MODELSED12}  with the following restrictions: $2000$K$<$T$_{\text{eff}}$<$5000$K, $3<\log g<6$, and $-1.0<$[Fe/H]$<-0.5$. The best fit from both models is T$_{\text{eff}}=3500\pm50$K, $\log g=4\pm0.25$ and [Fe/H] = $-0.5\pm0.25$. We note that the minor contamination from the background source in W1 and W2 is just about apparent when comparing the SED to the best fit model. We then used the {\tt isochrones} python package \citep{2015ascl.soft03010M}, with the Dartmouth model grid \citep{Do08} to obtain a mass estimate for l2 of $0.38\pm0.06M_{\sun}$.

Using the mass derived from l2's photometry we estimate its Einstein radius is $\Theta_{E}=4.56^{+0.35}_{-0.37}$\, mas. Consequently, we find a peak photometric amplification of the lens-source blend flux of $1.01^{+0.04}_{-0.01}$ or a change in magnitude of $10^{+40}_{-10}$ mmag. The Einstein time scale for this event is $7.7\pm0.6$ d. At the point l2 is still resolvable from the background source by there is a maximum astrometric shift due the major source image of $0.20^{+0.04}_{- 0.03}$\, mas. Over the event maximum when the lens and source are blended we predict a peak astrometric shift of the lens source light centroid of $0.04\pm0.01$\, mas.

This event was also independently predicted by K18 as event \#3. We find we are in good agreement with both K18's astrometric calculations, mass estimate for the lens, and consequently the predicted photometric and astrometric signals for this event. This event is not in B18's sample as both l2 and the background source fail B18's astrometric excess noise significance quality cuts.

\section{Observational Outlook}

\subsection{Astrometric Signal}

Both candidate 1 and 2 have relatively low astrometric deflection magnitudes, which will be challenging to detect. This is due to a combination of the specific event geometry and contaminating flux from the lens suppressing the astrometric signal when the lens and source are unresolved around the event maximum. For {\it Gaia}, an important predictor of the precision at which a deflection can be measured is the scan direction relative to the deflection direction. For measurements in which the deflection is aligned along {\sl Gaia}'s scan direction (AL) measurements will precise, whereas measurements aligned in the across scan direction (AC) will be considerably less precise \citep{F16}.

Recent simulations of the astrometric centroiding precision of {\it Gaia} have been carried out by \cite{R18}. For objects with G band magnitude $\sim$15, $\sim$17 and $\sim$18 Rybicki reports AL precisions $\sigma_{\text{AL}}\sim 0.1,0.5,0.7$\,mas, and AC precisions $\sigma_{\text{AC}}\sim3,20,60$\, mas (see \cite{R18} tables 1 and 2, for astrometric precision as a function of Johnson V mag, and B18 Fig 4. for the conversion to {\it Gaia} G).

In the case of candidate 1, the peak astrometric deflection of the $G\sim 18$ source at the point the lens and source are still resolvable is $\delta_{\text{r,max}}\sim 0.077$\,mas, which is < $\sigma_{\text{AL}}$. However, around event maximum, the $G\sim 17$ lens source blend will shift by $\delta_\text{max}\sim0.57$\,mas which is > $\sigma_{\text{AL}}$. This means candidate 1 should be borderline astrometrically detectable by {\it Gaia} around its maximum, but for only the most favorable scan directions.

We find the opposite case for candidate 2. When the lens and source are resolved $\delta_{\text{r,max}} > \sigma_{\text{AL}}$ as well as around closest approach, when they are not, we find the predicted astrometric shifts $\delta_{\text{max}}< \sigma_{AL}$. This means the tails of the candidate 2 event will only be astrometrically detectable by {\it Gaia} and only for the most favorable scan alignments. Overall this presents a pessimistic outlook for mass measurements of l1 and l2 solely derived from the astrometric deflection measurements by {\it Gaia}.

A better option may be to monitor the events with the Hubble Space Telescope (HST). Campaigns to measure the mass of single stars via astrometric microlensing with HST are already underway \citep{Kains2017}, and the mass of white dwarf Stein 2051 b was determined with $\sim 8$ per cent precision via astrometric microlensing with HST \citep{Sahu2017}. Although the amplitude of the deflection for both candidates 1 and 2 is small, so is the contrast ratio between the lens and source, which may allow better resolution and detection of the deflection with HST. 

\subsection{Photometric Signal}

For candidates 1 and 2, the predicted peak amplifications of the source lens blends are $\sim2$ and $\sim 1$ per cent corresponding to changes in magnitude of $\sim$ 20 and $\sim$ 10 mmag respectively. As pointed out by B18 for candidate 1, the signal could be boosted by a careful choice of the filter used. This is because l1 and the background source differ considerably in colour (see table \ref{tab:candevents}). 

Photometric precision between $\sim$ 1 and 10 mmag is typically achieved from ground based microlensing surveys \citep[e.g][]{U15,S15}, and $\sim$  mmag precision is routinely achieved from the ground in the studies of transiting exoplanets \cite[e.g.][]{G16,B18,D17}. Although our candidate events are on the fainter side with candidate 1 and 2 being $\sim$ 14 and 11 mag in 2MASS K band, photometric follow-up from the ground looks optimistic. 

\section{Simulated Mass Determination}

In this section, we investigate the efficiency of our method in determining the mass of the lens for our two candidate events. We generate synthetic light curve data according to eq.~ (\ref{eq:model}) for both events. Specifically for candidate 1, we take a draw of truth parameters from our estimate of the mass of l1 (a Gaussian centered on 0.11 and with standard deviation 0.01 $M_{\sun}$) and Gaussians centered at the {\it Gaia} DR2 mean and with a width of the DR2 measurement variance for the rest of the model parameters outlined in eq.~(\ref{eq:model}). These truth parameters are then used to generate 50 synthetic data points uniformly distributed around the event maximum. The data points are then scattered with independent Gaussian noise ($\epsilon_{i} \sim \mathcal{N}(0,1)\text{ mmag})$ and given error bars of $\sim$ 1 mmag photometric precision (see Fig.~\ref{fig:corner}, Top right).

We sample from the posterior distribution outlined in eq.~(\ref{eq:posteroir}) using the Affine Invariant Markov chain Monte Carlo (MCMC) Ensemble sampler, implemented by the {\tt EMCEE} package \citep{FM13}, with the priors outlined in table \ref{tab:priors} derived from DR2 quantities for the source and lens. We initialize 150 walkers in a small Gaussian ball around the maximum likelihood estimate and run a burn in of 150 steps. We then run the sampler for a further 150 steps keeping the final 100 steps.

Fig.~\ref{fig:corner} shows the marginal distributions from the posterior for one set of generated event data for the candidate 1 event. The generated data and model samples from the inference are also shown. Fig.~\ref{fig:corner} shows that for this particular set of generated data we were able to recover a mass for L1 of $M_{l}=0.11\pm0.01M_{\sun}$, which is consistent with the truth lens mass used to generate the data $M_{l,\text{true}}=0.111 M_{\sun}$ $\pm 10$ per cent. Fig.~\ref{fig:corner} also shows a degeneracy in relative proper motion of the source and lens ($\mu_{\alpha *,\text{rel}},\mu_{\delta,\text{rel}}$) and the lens mass $M_{l}$, which is related to the well-known hard degeneracy between the Einstein time and the lens mass in the light curve.

In order to estimate the average precision at which this method could determine the mass of l1, we run the inference 1000 times and stack the normalised marginal lens mass distributions, that is the samples of $(M_{l}-M_{l,\text{true}}) / M_{l,\text{true}}$, in each case. We repeat the same exercise for candidate 2. We handle the negative parallax measurement for the source star (see table \ref{tab:candevents}) by setting the source parallax to zero. The means the prior on $\varpi_{rel}$ and distribution used to generate the synthetic data is just  $\varpi_{\text{rel}}\sim \mathcal{N}(\varpi_{l},\sigma^{2}_{\varpi_{l}})$.

Fig.~\ref{fig:efficiency} shows the $(M_{l}-M_{l,\text{true}}) / M_{l,\text{true}}$ samples for the 1000 realisations of the inference for both the candidate 1 and candidate 2 events. For candidate 1 this suggests that the mass can be recovered to within 20 per cent (68 per cent confidence limit). This could be improved by choosing a filter to suppress the flux from the lens and increase the signal. 

For candidate 2, we see the recovered mass distribution is peaked at zero. However, there is a larger standard deviation of $\pm47$ per cent in the precision at which we can recover the mass of l2. Fig. \ref{fig:efficiency} also shows a truncation in the distribution of recovered normalised masses at -1.0. This is due to the uniform prior (see table \ref{tab:priors}) used for the lens mass in the inference forbidding unphysical negative lens masses. This relatively poor result compared with candidate 1 is mainly caused by two factors. First, the event geometry for candidate 2 results in a smaller signal-to-noise ratio. Secondly, the astrometry for both lens and background source are worse for candidate 2 than candidate 1. This results in less tight priors in modelling and hence a poorer constraint on the lens mass.

For both candidates, we have assumed a uniform sample of 50 data points around the peak amplification. If the sampling is not uniform, but has more measurements around the event maximum which we can predict to within a $\sim$ day, then the accuracy of the mass estimate will be improved. We also note that future {\it Gaia} data releases (DR3 expected $\sim$ 2020\footnote{\url{https://www.cosmos.esa.int/web/gaia/release}}) will provide better astrometric solutions and therefore will provide tighter priors in the modeling of the lightcurves. This will help to increase the level of precision on the inferred lens masses, especially for l2.

There are also some caveats. First, we have assumed a point mass and point lens, yet the lens could have a companion, which will make modelling significantly more challenging. Second, we have assumed mmag precision over the course of a couple of weeks. And finally, we have assumed that there is no intrinsic variability in the flux both from the source and lens. This could be of the same order of the microlensing signal - although in principle this could also be modelled.

\section{Conclusions}

The Gaia satellite is proving a tremendous resource for microlensing~\citep{Be02, Pr11}. Here, we have identified two high amplitude microlensing events with a systematic search through the {\sl Gaia} Data Release 2. Our focus here is to predict the time of closest approach between lens and source, which is the time of maximum amplification. With this in hand, the lensing event can be densely sampled using ground-based telescopes.

Our candidate event 1 has been reported before by B18. However, B18 seems to have assumed that the lens was an ordinary field M dwarf and as a result has significantly overestimated its mass. Our SED fits and the subsequent X-shooter spectrum that we have obtained are consistent with the lens being an M3 type ultra-subdwarf of mass $\approx 0.11 M_\odot$. This is a double-edged sword. On the one hand, mass estimates of cool subdwarfs from lensing are of greater scientific interest. On the other hand, the predicted Einstein radius is smaller than calculated by B18 because of the much lower mass. We predict that the time of maximum amplification is November 3rd 2019 ($\pm$1d).

The lens in our candidate event 2 is a metal-poor M dwarf with mass $\approx 0.38 M_\odot$. This is derived from SED fitting using {\sl Gaia}, VISTA and {\sl WISE} photometry. The epoch of maximum amplification occurs on June 3rd 2019 ($\pm$4d). This event was also found by K18.

Rather than follow these events up astrometrically (which would be challenging), we suggest that both are excellent candidates for dense sampling of the photometric lightcurves from the ground. This exploits the excellent {\sl Gaia} astrometric solution to provide priors for all the lightcurve model parameters (with the exception of the lens mass). With current {\it Gaia} DR2 data we have shown that this can provide the lens mass l1 to $\approx 20$ per cent and l2 to $\approx 47$ per cent. There is a opportunity for greater precision if the lightcurve sample is denser at maximum, photometric filters are chosen carefully, and improved astrometric data from future {\it Gaia} data releases is used. This provides a new way to extract microlensing masses for forthcoming events that can be predicted in the {\sl Gaia} data releases.

\section*{Acknowledgements}

PM would like to thank the Science and Technologies Research Council (STFC) for studentship funding.
This work has made use of data from the European Space Agency (ESA) mission \textit{Gaia}\footnote{\url{https://www.cosmos.esa.int/gaia}}, processed by the \textit{Gaia} Data Processing and Analysis Consortium (DPAC).\footnote{\url{https://www.cosmos.esa.int/web/gaia/dpac/consortium}} Funding for the DPAC has been provided by national institutions, in particular the institutions participating in the \textit{Gaia} Multilateral Agreement. This publication makes use of data products from the Two Micron All Sky Survey, which is a joint project of the University of Massachusetts and the Infrared Processing and Analysis Center/California Institute of Technology, funded by the National Aeronautics and Space Administration and the National Science Foundation. This publication makes use of VOSA, developed under the Spanish Virtual Observatory project supported from the Spanish MINECO through grant AyA2017-84089.



\bibliographystyle{mnras}
\bibliography{refs} 






\bsp	
\label{lastpage}
\end{document}